# Control Function Instrumental Variable Estimation of Nonlinear Causal Effect Models

**Zijian Guo**                                             ZIJGUO@WHARTON.UPENN.EDU
*Department of Statistics*
*University of Pennsylvania*
*Philadelphia,19104, USA*

**Dylan Small**                                            DSMALL@WHARTON.UPENN.EDU
*Department of Statistics*
*University of Pennsylvania*
*Philadelphia,19104, USA*

**Editor:**

## Abstract

The instrumental variable method consistently estimates the effect of a treatment when there is unmeasured confounding and a valid instrumental variable. A valid instrumental variable is a variable that is independent of unmeasured confounders and affects the treatment but does not have a direct effect on the outcome beyond its effect on the treatment. Two commonly used estimators for using an instrumental variable to estimate a treatment effect are the two stage least squares estimator and the control function estimator. For linear causal effect models, these two estimators are equivalent, but for nonlinear causal effect models, the estimators are different. We provide a systematic comparison of these two estimators for nonlinear causal effect models and develop an approach to combing the two estimators that generally performs better than either one alone. We show that the control function estimator is a two stage least squares estimator with an augmented set of instrumental variables. If these augmented instrumental variables are valid, then the control function estimator can be much more efficient than usual two stage least squares without the augmented instrumental variables while if the augmented instrumental variables are not valid, then the control function estimator may be inconsistent while the usual two stage least squares remains consistent. We apply the Hausman test to test whether the augmented instrumental variables are valid and construct a pretest estimator based on this test. The pretest estimator is shown to work well in a simulation study. An application to the effect of exposure to violence on time preference is considered.

## 1. Introduction

Randomized controlled studies are the gold standard to estimate the treatment effect. Unfortunately, randomized controlled studies are often not feasible because of cost or ethical constraints. When randomized studies are not feasible, observational studies provide an alternative source of data for estimating the treatment effect. Since treatments were not randomly assigned, a major concern in observational studies is the possible presence of un-measured confounders that affect both the treatment and the outcome. The instrumental variable method is designed to estimate the effects of treatments when there are unmeasured confounders. The method requires a valid instrumental variable which is a variable that (1) is associated with the treatment conditioning on the measured covariates; (2) is independent of the unmeasured confounders conditioning on the measured covariates; (3) has no direct effect on the outcome beyond its effect on the treatment. The definition of an instrumental variable is illustrated in Figure 1, where the arrows represent the causal relationship. The lack of an arrow between the instrumental variable and unmeasured confounding represents assumption (2); The lack of an arrow between the instrumental variable and outcome represents assumption (3). The existence of an arrow between the treatment and instruments represents assumption (1) when the instrumental variable causes the treatment. Note that assumption (1) is also satisfied when the instrumental variable is associated with the treatment, but does not cause the treatment, see Hernan & Robins (2006). See Holland (1988), Angrist et al. (1996), Tan (2006), Cheng et al. (2009), Brookhart et al. (2009), Baiocchi et al. (2014) and Imbens (2014) for good discussions of instrumental variable methods.

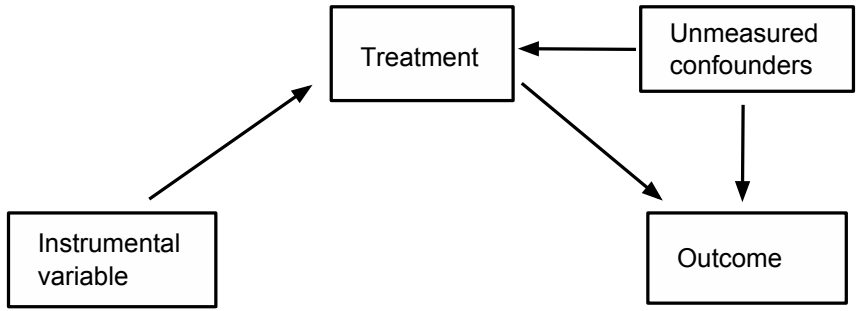

Figure 1: The directed acyclic graph for the relationship between the instrumental variable, treatment, unmeasured confounders and outcome.

Two commonly used methods for using an instrumental variable to estimate the treatment effect are the two stage least squares method and the control function method. Both of these methods are implemented in two stages. In the first stage of the two stage least squares method, we regress the treatment on the instruments and baseline covariates. In the second stage, the outcome is regressed on the predicted value from the first stage and baseline covariates and the coefficients of the second stage regression are taken as the two stage least squares estimates. Compared to the ordinary least squares method, the second stage of the two stage least squares method replaces the treatment with the corresponding predicted value from the first stage. The basic idea of the two stage least squares method is to extract variation in the treatment that is independent of the unmeasured confounders and use this variation to estimate the treatment effect. In contrast to two stage least squares, the control function method uses the instrumental variable to split the unmeasured confounders into two parts: one part that is correlated with the treatment and the other part that is uncorrelated with the treatment. The first stage of the control function method is similar to that of the two stage least square method. In the second stage of the control function method, the outcome is regressed on the treatment, baseline covariates and the residual of the first stage regression and the coefficients of the second stage regression are taken as the control function estimates. Intuitively, the residual of the first stage regression accounts for the unmeasured confounders. The method is called the control function method because the effect of the unmeasured confounders is controlled for via the inclusion of the control function (residual of the first stage regression) in the second stage regression. The control function method has been developed in econometrics (Heckman (1976); Ahn & Powell (2004); Andrews & Schafgans (1998); Blundell & Powell (2004); Imbens & Wooldridge (2007)) and also in biostatistics and health services research (Nagelkerke et al (2000) and Terza et al. (2008)), where the method has been called the two stage residual inclusion method.

When the treatment has a linear effect on the outcome, then the control function method and two stage least squares produce identical estimates. However, when the treatment has a nonlinear effect on the outcome, then the methods produce different estimates. Settings in which the treatment is thought to have a nonlinear effect on the outcome are common; examples include the effect of the concentration of a drug in a person's body on the body's response (Nedelman et al, 2007), the effect of education on earnings (Card, 1994), the effect of a bank's financial capital on its costs (Hughes & Mester, 1998) and the effect of industry lobbying on government tariff policy (Gawande & Bandyopadhyay, 2000).

Imbens & Wooldridge (2007) conjectured that the control function estimator, while less robust than two stage least squares, might be much more precise because it keeps the treatment variables in the second stage. However, Imbens and Wooldridge note that a systematic comparison had not been done between the two estimators. The goal of this paper is to provide such a systematic comparison of the control function method to two stage least squares for nonlinear causal effect models.

We illuminate the relationship between control function and two stage least squares estimators by showing that the control function estimator is equivalent to a two stage least squares estimator with an augmented set of instrumental variables. When these augmented instrumental variables are valid, the control function method is more efficient than usual two stage least squares. For example, in the setting (1) of Table 1, the control function estimator is considerably more efficient than two stage least squares, sometimes more than 10 times more efficient. However, when the augmented instrumental variables are invalid, the control function method is inconsistent. Thus, the key issue for deciding whether to use the control function estimator vs. usual two stage least squares is whether the augmented instrumental variables are valid. The validity of the augmented instrumental variables can be tested using the Hausman test (Hausman, 1978). We develop a pretest estimator based on this test and show that it performs well in simulation studies, being close to the control function estimator when the augmented instrumental variables are valid and close to two stage least squares when the augmented instrumental variable are invalid. The pretest estimator combines the strengths of the control function and two stage least squares estimators.

Our paper is organized as follows. In section 2, we describe the set up of our model and how to implement the two stage least squares method and the control function method. In section 3, we describe how the control function method is equivalent to two stage least squares with an augmented set of instrumental variables. In section 4, we describe the Hausman test for the validity of the augmented instrumental variables and formulate a pretest estimator. In section 5 , we present simulation studies comparing the control function method to the usual two stage least squares method. In section 6, we apply our methods to a study of the effect of exposure of violence on time preference. In section 7, we present discussions about more generalized models. In section 8, we conclude the paper. The proof of theorems, more simulation studies and two additional data analysis are presented in the Appendix.

## 2. Set up of model and description of methods

### 2.1 Set up of model

In this paper, our focus is on a model where the outcome variable is a non-linear function of the treatment variable. Let $y_1$ denote the outcome variable and $y_2$ denote the treatment variable. Let $z_1$ denote a vector of measured pre-treatment covariates and $z_2$ denote a vector of instrumental variables and $z = (z_1, z_2)$. In the following discussion, we assume both $z_1$ and $z_2$ are one dimensional. In section 3.4, we will extend the method to allow $z_1$ and $z_2$ to be vectors.

For defining the causal effect of the treatment, we use the potential outcome approach (Rubin (1974) and Neyman (1923)). Let $y_1^{(y_2^*)}$ denote the outcome that would be observed if the unit is assigned treatment level $y_2 = y_2^*$. An additive, non-linear causal effect model for the potential outcomes (similar to the linear causal effect model in Holland (1988)) is

$$y_1^{(y_2^*)} = y_1^{(0)} + \beta_1 g_1(y_2^*) + \beta_2 g_2(y_2^*) + \cdots + \beta_k g_k(y_2^*), \tag{1}$$

where $g_1(y_2^*) = y_2^*$ and $g_1, \cdots, g_k$ are linearly independent functions. Since $g_1$ is linear, $g_2, \cdots, g_k$ are non-linear functions. The causal effect of increasing $y_2$ from $y_2^*$ to $y_2^* + 1$ is

$$\left(\beta_1(y_2^* + 1) + \beta_2 g_2(y_2^* + 1) + \cdots + \beta_k g_k(y_2^* + 1)\right) - \left(\beta_1 y_2^* + \beta_2 g_2(y_2^*) + \cdots + \beta_k g_k(y_2^*)\right).$$

We assume that the pretreatment covariate $z_1$ has a linear effect on potential outcomes, $E\left(y_1^{(0)} | z_1\right) = \beta_0 + \beta_{k+1} z_1$ and denote the residual by $u_1 = y_1^{(0)} - E(y_1^{(0)} | z_1)$. The model for the observed data is

$$y_1 = \beta_0 + \beta_1 g_1(y_2) + \beta_2 g_2(y_2) + \cdots + \beta_k g_k(y_2) + \beta_{k+1} z_1 + u_1, \tag{2}$$

where $y_2$ is the observed value of the treatment variable and $g_1(y_2) = y_2$. For identifiability, we assume that $g_i(0) = 0$ for $2 \leq i \leq k$. We also assume that $y_2$'s expectation given $z_1$ and $z_2$ is linear in parameters but possibly nonlinear in $z_1$ and $z_2$, i.e.,

$$y_2 = \alpha_0 + \alpha_1^T \mathbf{J}(z_1) + \alpha_2^T \mathbf{H}(z_2) + v_2, \tag{3}$$

where $\alpha_2 \neq \mathbf{0}$, $z_1$ and $z_2$ are independent of $u_1$ and $v_2$, $u_1$ and $v_2$ are potentially correlated, and $\mathbf{H} = (z_2, h_2(z_2), \cdots, h_k(z_2))$ is a known vector of linearly independent functions of $z_2$. $\mathbf{J}$ is a known vector of functions, which can be a nonlinear function, for simplicity of notation, we will assume a linear effect of $z_1$ henceforth,

$$y_2 = \alpha_0 + \alpha_1 z_1 + \alpha_2^T \mathbf{H}(z_2) + v_2, \tag{4}$$

We say that $I(z_2) = (z_2, h_2(z_2) \cdots, h_k(z_2))$ are valid instrumental variables if $I(z_2)$ satisfies the following two assumptions (Stock (2002), White (1984)(p.8) and Wooldridge (2010)(p.99)):

**Assumption 1** (***Relevance***) *The instruments are correlated with the endogenous variables* $W = (g_1(y_2), \cdots, g_k(y_2))$ *given* $z_1$, *that is,* $E\left(W I(z_2)^T | z_1\right)$ *is of full column rank.*

**Assumption 2** (***Exogeneity***) *The instruments are uncorrelated with the error in* (2), *that is,*

$$E(I(z_2)u_1) = 0.$$

To ensure assumption 1, we need to have at least $k$ valid instrument variables for $g_1(y_2), \cdots, g_k(y_2)$. This requires that the instrument $z_2$ has at least $k$ different values. Otherwise, we cannot construct $k$ linearly independent functions of $z_2$ to be instruments for $g_1(y_2), \cdots, g_k(y_2)$.

**Remark 1** *In econometric terminology,* $y_2$ *is called an endogenous variable,* $z_1$ *are called included exogenous variables and* $I(z_2) = (z_2, h_2(z_2) \cdots, h_k(z_2))$ *is called an excluded exogenous variable. Sometimes* $z_1, z_2, h_2(z_2), \cdots, h_k(z_2)$ *together are referred to as the instrumental variables (i.e., both the included and excluded exogenous variables), but we will just refer to* $I(z_2) = (z_2, h_2(z_2) \cdots, h_k(z_2))$ *(the excluded exogenous variable) as the instruments. Assumption 1 is called the rank condition in the econometric literature while* $r \geq k$ *is called the order condition. As discussed in Wooldridge (2010)(p.99), the order condition is necessary for the rank condition.*

**Remark 2** *Sufficient conditions for Assumption 2 to hold are that (i) $I$ is independent of the potential outcome $y_1^{(0)}$ after controlling for the measured confounders $z_1$, that is $I$ is independent of unmeasured confounders and (ii) $I$ has no direct effect on $y_1$ and only affects $y_1$ through $y_2$. These conditions are discussed by Holland (1988).*

## 2.2 Description of two stage least squares method

In the following, we will describe the two stage least squares (2SLS) method briefly. Let $y \sim x_1 + x_2$ denote linear regression of $y$ on $x_1$, $x_2$ and an intercept. Following Kelejian (1971), the usual two stage least squares estimator is introduced as Algorithm 1.

We will refer to this as the usual two stage least squares method as we will show in section 3 that the control function method can be viewed as a two stage least squares method with an augmented set of instruments.

Let $L(y|x)$ denote the best linear projection $\gamma^T x$ of $y$ onto $x$ where $\gamma = \arg\min_\tau E\left((y - \tau^T x)^2\right)$. Let $Z = \{1, z_1, z_2, h_2(z_2), \cdots, h_k(z_2)\}$ denote the exogenous variables. The linear projection of $y_1$ onto $1, L\left(g_1(y_2)|Z\right), \cdots, L\left(g_k(y_2)|Z\right), z_1$ is

$$L\left(y_1 | 1, L\left(g_1(y_2)|Z\right), \cdots, L\left(g_k(y_2)|Z\right), z_1\right) = \beta_0 + \sum_{i=1}^{k} \beta_i L\left(g_i(y_2)|Z\right) + \beta_{k+1}z_1.$$

Thus, if we know $L = \{L\left(g_1(y_2)|Z\right), \cdots, L\left(g_k(y_2)|Z\right)\}$, then the least squares estimate of $y$ on $1, L\left(g_1(y_2)|Z\right), \cdots, L\left(g_k(y_2)|Z\right), z_1$ will provide an unbiased estimate of $\beta$. Even though $L$ is unknown, we can obtain a consistent estimate of $L$. Substituting this estimate into the second stage regression provides a consistent estimate of $\beta$ under regularity conditions. See White (1984)(p21, p32) for the details.

---

**Algorithm 1** Two stage least squares estimator (2SLS)

---

**Input:** i.i.d observations of pre-treatment covariates $z_1$, instrumental variables $(z_2, h_2(z_2) \cdots, h_k(z_2))$, the treatment variable $y_2$ and outcome variable $y_1$.

**Output:** The two stage least squares estimator $\widehat{\beta}^{\text{2SLS}}$ of treatment effects $\beta = (\beta_1, \beta_2, \cdots, \beta_k)$ in (2).

1: Implement the regression

$$y_2 \sim z_1 + z_2 + h_2(z_2) + \cdots + h_k(z_2),$$
$$\vdots \qquad\qquad\qquad\qquad (5)$$
$$g_k(y_2) \sim z_1 + z_2 + h_2(z_2) + \cdots + h_k(z_2),$$

and obtain the corresponding predicted values as $\left( \widehat{y_2}, \widehat{g_2(y_2)}, \cdots, \widehat{g_k(y_2)} \right)$.

2: Implement the regression:

$$y_1 \sim \widehat{y_2} + \cdots + \widehat{g_k(y_2)} + z_1. \qquad (6)$$

The two stage least squares estimates are the coefficient estimates on $y_2, g_2(y_2), \cdots, g_k(y_2)$ from (6).

---

## 2.3 Description of the control function method

We introduce the control function (CF) method in Algorithm 2.

---

**Algorithm 2** Control function estimator (CF)

---

**Input:** i.i.d observations of pre-treatment covariates $z_1$, instrumental variables $(z_2, h_2(z_2) \cdots, h_k(z_2))$, the treatment variable $y_2$ and outcome variable $y_1$.

**Output:** The control function estimator $\widehat{\beta}^{\text{CF}}$ of treatment effects $\beta = (\beta_1, \beta_2, \cdots, \beta_k)$ in (2).

1: Implement the regression

$$y_2 \sim z_1 + z_2 + h_2(z_2) + \cdots + h_k(z_2) \qquad (7)$$

and obtain the predicted value $\widehat{y_2}$ and the residual $e_1 = y_2 - \widehat{y_2}$.

2: Implement the regression

$$y_1 \sim y_2 + \cdots + g_k(y_2) + z_1 + e_1. \qquad (8)$$

The control function estimates are the coefficient estimates on $y_2, g_2(y_2), \cdots, g_k(y_2)$ from the above regression (8).

---

## 2.4 Additional assumptions of control function

The consistency of control function estimators requires stronger assumptions than $I(z_2)$ satisfying Assumptions 1 and 2, that is, stronger than $I(z_2)$ being valid instrumental variables.

The control function method is consistent under the following additional assumptions in addition to Assumptions 1 and 2 (Imbens & Wooldridge , 2007):

**Assumption 3** $u_1$ and $v_2$ are independent of $z = (z_1, z_2)$.

**Assumption 4** $E(u_1|v_2) = L(u_1|v_2)$ where $L(u_1|v_2) = \rho v_2$ and $\rho = \frac{E(u_1 v_2)}{E(v_2^2)}$.

**Remark 3** We only assume $z = (z_1, I(z_2))$ is uncorrelated with $u_1$ in Assumption 2. Two stage least squares does not make assumptions about $v_2$ but the control function method relies on assumptions about $v_2$. If $(u_1, v_2)$ follow bivariate normal distribution, Assumption 4 is satisfied automatically.

Model (29) is an example where $I(z_2) = (z_2, z_2^2)$ are valid instrumental variables that satisfy Assumptions 1 and 2 but not Assumption 3 and 4. The simulation results in Table 1 show that the control function estimate for model (29) is biased.

Assumptions 3 and 4 are sufficient but not necessary conditions for the control function method to be consistent. To investigate the consistency of the control function method, we can express (2) as

$$y_1 = \beta_0 + \beta_1 g_1(y_2) + \beta_2 g_2(y_2) + \cdots + \beta_k g_k(y_2) + \beta_{k+1} z_1 + \rho v_1 + e, \qquad (9)$$

where $e$ is $u_1 - \rho v_2$. Under some regularity conditions (Wooldridge (2010) p.56), one sufficient condition for the control function method to be consistent is that the new error $e$ in (9) is uncorrelated with $g_i(y_2)$:

$$E(g_i(y_2)e) = 0 \quad \text{for} \quad i = 1, \cdots, k. \qquad (10)$$

Assumptions 1,2, 3 and 4 lead to $E(f(y_2)e) = 0$ for all $f$, which is stronger than the sufficient condition (10). In Section 4, we develop a test for the consistency of the control function method that can help to decide whether the control function estimator should be used.

## 3. Relation between control function and two stage least squares

### 3.1 Loss of information for the control function method and comparison to two stage least squares

To better understand the control function method, we will focus on the case $k = 2$

$$y_1 = \beta_0 + \beta_1 y_2 + \beta_2 g_2(y_2) + \beta_3 z_1 + u_1;$$

$$y_2 = \alpha_0 + \alpha_1 z_1 + \alpha_2 z_2 + \alpha_3 h_2(z_2) + v_2.$$

For example, a common model has $g_2(y_2) = y_2^2$ and $h_2(z_2) = z_2^2$. Two stage least squares regresses $y_1$ on $\widehat{y_2} = LS(y_2|1, z_1, z_2, h_2(z_2))$, $\widehat{g_2(y_2)} = LS(g_2(y_2)|1, z_1, z_2, h_2(z_2))$ and $z_1$ where $LS(A|B)$ denotes the least square estimate of the linear projection of $A$ onto span $(B)$. The control function estimator regresses $y_1$ on $y_2, g_2(y_2), z_1$ and $e_1 = y_2 - \widehat{y_2}$. Intuitively, it seems that the control function method is better than two stage least squares since it keeps $y_2$ and $g_2(y_2)$ in the second stage regression rather than losing information by replacing $y_2$ and $g_2(y_2)$ by projections as two stage least squares does. However, the control function method also loses information as explained in the following theorem.

**Theorem 1** *The following regression*

$$y_1 \sim z_1 + \widetilde{y_2} + \widetilde{g_2(y_2)},$$

*will have the same estimates of the corresponding coefficients as the second stage regression of the control function method:*

$$y_1 \sim z_1 + y_2 + g_2(y_2) + e_1,$$

*where $e_1$ is the residual of the regression $y_2 \sim z_1 + z_2 + h_2(z_2)$ and $\widetilde{y_2}$ is the residual of the regression $y_2 \sim e_1$ and $\widetilde{g_2(y_2)}$ is the residual of the regression $g_2(y_2) \sim e_1$. By the property of OLS, $\widetilde{y_2}$ differs from $\hat{y}_2$ only by a constant.*

**Proof** In appendix *B*. ∎

From Theorem 1, the control function method can be viewed as a two stage estimator similar to two stage least squares: in the first stage, the exogenous part of the endogenous variables, $\widetilde{y_2}$ and $\widetilde{g_2(y_2)}$, are obtained by taking residuals from regressing $y_2$ and $g_2(y_2)$ on the variable $e_1$ respectively and in the second stage, the endogenous variables in the regression are replaced by the exogenous parts.

Although both the control function method and two stage least squares lose information by projecting $g_2(y_2)$ and $y_2$ to subspaces, the control function method does preserve more information. Let $(\text{span}\{e_1\})^{\perp}$ denote the orthogonal complement of the space spanned by the vector $e_1$ and span$\{1, z_1, z_2, h_2(z_2)\}$ denote the space spanned by the vectors $1, z_1, z_2, h_2(z_2)$. The control function method projects $g_2(y_2)$ and $y_2$ to $(\text{span}\{e_1\})^{\perp}$ while two stage least squares projects $g_2(y_2)$ and $y_2$ to span$\{1, z_1, z_2, h_2(z_2)\}$, which is a subspace of $(\text{span}\{e_1\})^{\perp}$. We will now show that the control function method is equivalent to two stage least squares with an augmented set of instrumental variables.

### 3.2 Equivalence of control function to two stage least squares with an augmented set of instrumental variables

In section 3.1, we showed that the first stage projection space of the control function estimator is larger than that of the two stage least squares estimator. One natural question is what is the extra part of the control function first stage projection space. The following theorem explains the extra part.

**Theorem 2** *The following two regressions give the same estimates of corresponding coefficients*

$$g_2(y_2) \sim z_1 + z_2 + h_2(z_2);$$
$$\widetilde{g_2(y_2)} \sim z_1 + z_2 + h_2(z_2).$$

**Proof** In appendix *B*. ∎

Theorem 2 implies that regressing the control function first stage estimate of $g_2(y_2)$, $\widetilde{g_2(y_2)}$ on $\{1, z_1, z_2, h_2(z_2)\}$ produces the same predicted value for $g_2(y_2)$ as that of the first stage of two stage least squares. Consequently, the control function method first stage estimate $\widetilde{g_2(y_2)}$ can be decomposed into two orthogonal parts: one part is the same with the two stage least squares first stage estimate and the other part is denoted as $error.iv$. Letting $\gamma_0, \gamma_1, \gamma_2, \gamma_3$ denote the coefficients in $g_2(y_2) \sim z_1 + z_2 + h_2(z_2)$, we define

$$error.iv = \widetilde{g_2(y_2)} - (\gamma_0 + \gamma_1 z_1 + \gamma_2 z_2 + \gamma_3 h_2(z_2)). \tag{11}$$

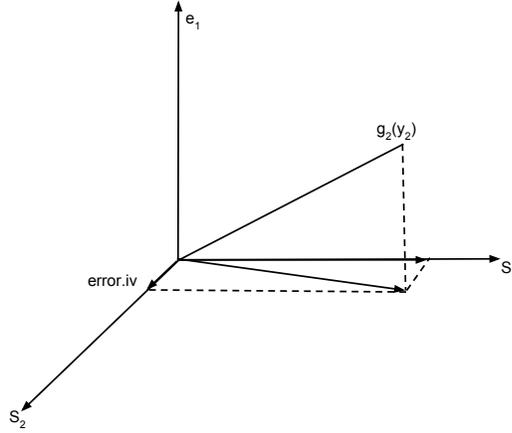

Figure 2: Projection and Decomposition of the control function Estimator. $S_1 = \text{span}\{1, z_1, z_2, h_2(z_2)\}$ and $S_2 = \text{span}\{error.iv\}$ is defined as the orthogonal complement of $S_1$ in the subspace $\text{span}\{e_1\}^{\perp}$.

The decomposition is shown in Figure 2. Let $S_1 = \text{span}\{1, z_1, z_2, h_2(z_2)\}$ and $S_2 = \text{span}\{error.iv\}$. The projection of $g_2(y_2)$ to the space $S_1 \cup S_2$ is the control function first stage predicted value $\widetilde{g_2(y_2)}$ while the projection of $g_2(y_2)$ to the space $S_1$ is the two stage least squares first stage predicted value $\widehat{g_2(y_2)}$. The projection along the direction of $error.iv$ represents the extra part of the control function projection.

Since $\widetilde{g_2(y_2)}$ is the residual of the regression $g_2(y_2) \sim e_1$, the first stage regression of control function estimation can be written as

$$
\begin{aligned}
g_2(y_2) &= \gamma_c + \gamma e_1 + \widetilde{g_2(y_2)} \\
&= \gamma_c + \gamma e_1 + \gamma_0 + \gamma_1 z_1 + \gamma_2 z_2 + \gamma_3 h_2(z_2) + error.iv
\end{aligned} \tag{12}
$$

where the column vector $e_1$ is orthogonal to the column vectors $z_2, h_2(z_2), error.iv$ and $error.iv$ is orthogonal to $z_1$, $z_2$ and $h_2(z_2)$ by the property of OLS. The following theorem shows the equivalence of the control function estimator to the two stage least squares method with an augmented set of instruments.

**Theorem 3** *The control function estimator with instruments $z_1, z_2, h_2(z_2)$ in the first stage is equivalent to the two stage least squares estimator with instruments $z_1, z_2, h_2(z_2), error.iv$ in the first stage.*

**Proof**  In appendix *B*.  ∎

We now show that if Assumptions 3 and 4 do hold, the extra instrument $error.iv$ is a valid instrumental variable, that is, $error.iv$ satisfies Assumptions 1 and 2. Recall the definition $e = u_1 - \rho v_2$. By Assumptions 3 and 4, $E(g_2(y_2)e) = 0$ and the population version of $error.iv$ is

$$g_2(y_2) - L(g_2(y_2)|v_2, z_1, z_2, h_2(z_2))$$

where $L$ means the linear projection. Since $v_2$ and $z_1, z_2, h_2(z_2)$ are uncorrelated with $e$,

$$E(error.iv \times e) = E((g_2(y_2) - L(g_2(y_2)|v_2, z_1,, h_2(z_2)))e) = 0.$$

### 3.3 Equivalence of control function with two stage least squares with augmented instruments for general nonlinear model

In the last section, we showed the equivalence of control function with two stage least squares for $k = 2$. In this section, we will show the results hold for $k \geq 3$ and present an algorithm to find the augmented set of instrumental variables for the general model. This will help us understand the control function estimator's properties since two stage least squares' properties are well understood (See Remarks 2 and section 3.5 below).

We will first consider $k = 3$ in our general model (2) and then it will be straightforward to generalize to $k > 3$. The model for $k = 3$ is

$$y_1 = \beta_0 + \beta_1 y_2 + \beta_2 g_2(y_2) + \beta_3 g_3(y_2) + \beta_4 z_1 + u_1.$$

Let $I = \{z_2, h_2(z_2), h_3(z_2)\}$ be valid instrumental variables for $y_2, g_2(y_2), g_3(y_2)$. As in the case of $k = 2$, we can find the extra instrument $error.iv_1$ for $g_2(y_2)$. If we do the same thing for $g_3(y_2)$, we will obtain the extra instrumental variable, $error.iv_2.prime$. However, there exists the problem that the error instrument $error.iv_2.prime$ is not orthogonal to $error.iv_1$. This will make $error.iv_2.prime$ have a non-zero coefficient when we treat $error.iv_2.prime$ as one of the instrumental variables for $g_2(y_2)$. The non-zero coefficient will lead to the estimate of $g_2(y_2)$ being different from what is used in the second stage of control function. To remove the correlation, we do the regression

$$error.iv_2.prime \sim error.iv_1,$$

and treat the residual of this regression as $error.iv_2$.

**Theorem 4** *If we regress $y_2$ on $z_1, z_2, h_2(z_2), h_3(z_2)$ in the first stage, then the control function estimator is equivalent to two stage least squares with the augmented set of instrumental variables $z_1, z_2, h_2(z_2), h_3(z_2), error.iv_1, error.iv_2$.*

**Proof** In appendix B. ■

**Remark 4** *Since the control function method is the same as two stage least squares with an augmented set of instrumental variables, we can calculate the variance of control function estimators with the help of the known variance formula for two stage least squares (Wooldridge (2010), page 102).*

This generalizes to the general $k > 3$ model. The following is an algorithm to identify the augmented set of instrumental variables for which the control function estimator is equivalent to two stage least squares with the augmented set of instrumental variables. Recall that $e_1$ denotes the residual of the regression $y_2 \sim z_1 + z_2 + h_2(z_2) + h_3(z_2)$ and $\widehat{g_i(y_2)} = resid(g_i(y_2) \sim e_1)$. Set the control function projection space $S_1 = \text{span}\{z_1, z_2, h_2(z_2), h_3(z_2)\}$ for $y_2$. Regress $\widehat{g_2(y_2)}$ on the space $S_1$ to obtain the residual $error.iv_1$ and set the control function projection space $S_2 = S_1 \oplus \text{span}\{error.iv_1\}$, where $\oplus$ denotes the direct sum of two subspaces. For $g_j(y_2), j = 3, \cdots, k$, regress $\widehat{g_j(y_2)}$ on $S_{j-1}$ to obtain the extra instrument $error.iv_j$, and set the control function projection space $S_j = S_{j-1} \oplus \text{span}\{error.iv_j\}$. This algorithm will produce the corresponding set of instrumental variables for the control function method. Each nonlinear function of endogenous variables will contribute one extra instrumental variable.

We can generalize Theorem 4 to models with $k > 3$.

**Theorem 5** *If we regress $y_2$ on $z_1, z_2, h_2(z_2), \cdots, h_k(z_2)$ in the first stage, then the control function estimator is equivalent to two stage least squares with the augmented set of instrumental variables $z_1, z_2, h_2(z_2), \cdots, h_k(z_2), error.iv_1, error.iv_2, \cdots, error.iv_{k-1}$.*

The proof of Theorem 4 applies to this generalized theorem.

### 3.4 General nonlinear model of multidimensional instruments $\mathbf{z}_2$

In this section, we extend our results to allow $\mathbf{z}_1$ and $\mathbf{z}_2$ to be vectors. In the vector case, the first stage model is extended as

$$y_2 = \alpha_0 + \alpha_1^T \mathbf{J}(\mathbf{z}_1) + \alpha_2^T \mathbf{H}(\mathbf{z}_2) + v_2, \tag{13}$$

where $\mathbf{J}$ is a known vector of functions of $\mathbf{z}_1$ and $\mathbf{H} = (\mathbf{z}_2, h_2(\mathbf{z}_2), \cdots, h_k(\mathbf{z}_2))$ is a known vector of linearly independent functions of $\mathbf{z}_2$.

The first stage of two stage least squares is extended as

$$
\begin{aligned}
y_2 &\sim \mathbf{J}(\mathbf{z}_1) + \mathbf{z}_2 + h_2(\mathbf{z}_2) + \cdots + h_k(\mathbf{z}_2), \\
&\vdots \\
g_k(y_2) &\sim \mathbf{J}(\mathbf{z}_1) + \mathbf{z}_2 + h_2(\mathbf{z}_2) + \cdots + h_k(\mathbf{z}_2),
\end{aligned}
\tag{14}
$$

where the notation $y \sim \mathbf{x}$, where $\mathbf{x} = (\mathbf{x}_1, \mathbf{x}_2, \cdots, \mathbf{x}_s)$ denotes linear regression of $y$ on $\mathbf{x}_1, \mathbf{x}_2, \cdots, \mathbf{x}_s$ and an intercept. From these k first stage regression, we obtain the predicted values of $k$ first stage regressions as $\left(\widehat{y_2}, \widehat{g_2(y_2)}, \cdots, \widehat{g_k(y_2)}\right)$. In the second stage, we do the following regression:

$$y_1 \sim \widehat{y_2} + \cdots + \widehat{g_k(y_2)} + \mathbf{z}_1. \tag{15}$$

The first stage of the control function method is

$$
\begin{aligned}
y_2 &\sim \mathbf{J}(\mathbf{z}_1) + \mathbf{z}_2 + h_2(\mathbf{z}_2) + \cdots + h_k(\mathbf{z}_2); \\
e_1 &= y_2 - \widehat{y_2}.
\end{aligned}
\tag{16}
$$

In the second stage, we do the following linear regression:

$$y_1 \sim y_2 + \cdots + g_k(y_2) + \mathbf{z}_1 + e_1. \tag{17}$$

The generalization of Theorem 4 and Theorem 5 is as follows:

**Theorem 6** *If we regress $y_2$ on $\mathbf{J}(\mathbf{z}_1), \mathbf{z}_2, h_2(\mathbf{z}_2), \cdots, h_k(\mathbf{z}_2)$ in the first stage, then the control function estimator is equivalent to two stage least squares with the augmented set of instrumental variables $\mathbf{J}(\mathbf{z}_1), \mathbf{z}_2, h_2(\mathbf{z}_2), \cdots, h_k(\mathbf{z}_2), error.iv_1, error.iv_2, \cdots, error.iv_{k-1}$.*

The proof of Theorem 4 applies to this generalized theorem.

### 3.5 Control function bias formula

To ensure consistency of the control function estimation, the Assumptions 1 and 2 of $z_2, h_2(z_2), \cdots, h_k(z_2)$ being valid instruments are not enough and extra assumptions are needed, such as Assumptions 3 and 4. We will present a formula for the asymptotic bias of the control function method when the extra assumptions of control function are not satisfied. Since we have shown that the control function estimator is a two stage least squares estimator with augmented instrumental variables (Sections 3.2 and 3.3), we could make use of the bias formula of two stage least squares with invalid instrumental variables given in Small (2007).

Let $Y_N$ denote the vector of the outcome variable $y_1$, $W_N$ denote the matrix whose columns are the endogenous variable $y_2, g_2(y_2), \cdots, g_k(y_2)$, $X_N$ denote the matrix whose columns are the included exogenous variables $z_1$ and $Z_N$ denote the matrix whose columns are the instrumental variables $z_2, h_2(z_2), \cdots, h_k(z_2)$ for $W_N$. Let $C_N = [W_N, X_N]$ and $D_N = [Z_N, X_N]$. In the second stage, we will replace $C_N$ by the first stage predicted values $\widehat{C_N} = D_N \left(D_N^T D_N\right)^{-1} D_N^T C_N$. The two stage least squares estimator is $\left(\widehat{C_N}^T \widehat{C_N}\right)^{-1} \widehat{C_N}^T Y_N$. When the instrumental variables are not valid, there exists asymptotic bias for the estimator given in the second stage. For the model

$$y_1 = \beta_0 + \beta_1 g_1(y_2) + \cdots + \beta_k g_k(y_2) + \beta_{k+1} z_1 + u_1 \tag{18}$$

we could write $u_1$ as

$$u_1 = LP(u_1|Z) + u_1 - LP(u_1|Z) \tag{19}$$

where $Z$ denote the instrumental variables and and $u_1 - LP(u_1|Z)$ is uncorrelated with $Z$ by the property of OLS. Plugging (19) into (18):

$$y_1 = \beta_0 + \beta_1 y_2 + \beta_2 y_2^2 + \beta_3 z_1 + LP(u_1|Z) + u_1 - LP(u_1|Z)$$

$$y_1 - LP(u_1|Z) = \beta_0 + \beta_1 y_2 + \beta_2 y_2^2 + \beta_3 z_1 + u_1 - LP(u_1|Z) \tag{20}$$

For (20), we could make use of the two stage least squares method with the outcome variable $y_1 - LP(u_1|Z)$ to obtain the consistent estimate $\left(\widehat{C_N}^T \widehat{C_N}\right)^{-1} \widehat{C_N}^T \left[Y_N - LP(U_N|Z_N)\right]$, where $U_N$ denotes the vector of error $u_1$ in the second stage. Hence a consistent estimator of the asymptotic bias from using the control function method is

$$\left(\widehat{C_N}^T \widehat{C_N}\right)^{-1} \widehat{C_N}^T LP\left(U_N | Z_N\right) \tag{21}$$

The control function bias formula (21) shows that in order for the control function estimator to be asymptotically unbiased, all augmented instruments are required to be uncorrelated with the error of the second stage. It is possible that the augmented instruments are correlated with the second stage error even when $Z_N$ are valid instrumental variables. See the simulation study in section 5.3 for an example. In Section 4, we develop a test for whether the augmented instruments are correlated with the error of the second stage.

### 3.6 Comparison between usual two stage least squares and control function

In this section, we compare usual two stage least squares (i.e., without any augmented instruments) to the control function method under the assumptions 1 and 2 of valid instruments and the additional assumptions 3 and 4 of the control function method. For $k = 3$, $X = (y_2, g_2(y_2), g_3(y_2))$ denote the endogenous variables and $Z_1 = (z_1, z_2, h_2(z_2), h_3(z_2))$ denote the instrumental variables. $Z_2 = (error.iv_1, error.iv_2)$ denote the augmented instruments and $\widetilde{X} = (e_1, \delta_0 e_1 + error.iv_1, \delta_1 e_1 + \delta_2 error.iv_1 + error.iv_2)$ denote the residual of regressing $X$ on $Z_1$. According to White (1984), the asymptotic variance of a two stage least squares estimator will not increase when adding extra valid instruments and will decrease as long as the extra valid instruments are correlated with the endogenous variables given the current instruments. In our setting, adding $Z_2$ to the list of instrumental variables will decrease the asymptotic variance if $Z_2$ is correlated with $\widetilde{X}$. Since $E(e_1 \times error.iv) = 0$, we have $E(error.iv_1 \widetilde{X}) = \left(0, E(error.iv_1^2), \delta_2 E(error.iv_1^2)\right)$ and $E(error.iv_2 \widetilde{X}) = \left(0, 0, E(error.iv_1^2)\right)$. Assume

$$g_2(y_2) \text{ is not a linear combination of } \quad 1, e_1, z_1, z_2 \text{ and } h_2(z_2) \quad \text{almost surely,} \tag{22}$$

we have $E(error.iv_1^2) > 0$ and $E(error.iv_1 \widetilde{X}) \neq 0$ and $E(error.iv_2 \widetilde{X}) \neq 0$. Hence, it follows that the control function estimator is strictly better than two stage least squares under the assumptions 1 and 2 of valid instruments, the additional assumptions 3 and 4 of the control function method and the functional assumption (22). Since $e_1$ is a linear function of $y_2$, the functional assumption (22) is saying that the nonlinear functional $g_2(y_2)$ cannot be expressed as a linear combination of linear functions of $y_2$ and the other variables $1, z_1, z_2, h_2(z_2)$. For example, assuming the first stage model (13) and $g_2(y_2) = y_2^2$ and $h_2(z_2) = z_2^2$, the functional assumption (22) is automatically satisfied.

We have shown that the control function estimator is equivalent to the two stage least squares estimator with an augmented set of instrumental variables and certain assumptions will make the control function estimator consistent and more efficient than the two stage least squares estimator. Just as we can view the control function estimator as one type of two stage least squares estimator, we can view the usual two stage least squares estimator without the augmented instruments as one type of control function estimator. For $k = 3$, if we include $e_1, e_2$ and $e_3$, which are the residuals of the regression $y_2 \sim z_1 + z_2 + h_2(z_2) + h_3(z_2)$, $g_2(y_2) \sim z_1 + z_2 + h_2(z_2) + h_3(z_2)$ and $g_3(y_2) \sim z_1 + z_2 + h_2(z_2) + h_3(z_2)$ respectively, in the second stage regression, this control function estimator is the same as the usual two stage least squares estimator without the augmented instruments.

## 4. A pretest estimator

We will explain how to test the validity of the augmented instrumental variables the control function method uses for $k = 2$; it is straightforward to extend the test to the more general model. Recall the instruments for the usual two stage least squares estimator: $Z_1 = (1, z_1, z_2, h_2(z_2))$; As we have shown (Sections 3.2 and 3.3), the control function estimator is a two-stage estimator with the instruments $Z = (Z_1, error.iv)$, where $error.iv$ is the augmented instrumental variable. We use $\widehat{\beta}^{\mathrm{2SLS}}$ to denote the usual two least squares estimator with instruments $Z_1$ and use $\widehat{\beta}^{\mathrm{CF}}$ to denote the two stage least squares estimator with the augmented set of instruments $Z$. Under the null hypothesis that the augmented instrumental variable $error.iv$ is valid, the two estimators are consistent and $\widehat{\beta}^{\mathrm{CF}}$ is efficient. Under the alternative hypothesis that the augmented instrumental variable $error.iv$ is invalid, $\widehat{\beta}^{\mathrm{2SLS}}$ is consistent while $\widehat{\beta}^{\mathrm{CF}}$ is not consistent.

The Hausman test (Hausman (1978)) provides a test for the validity of the augmented instrumental variable $error.iv$. The test statistic $H(\widehat{\beta}^{\mathrm{CF}}, \widehat{\beta}^{\mathrm{2SLS}})$ is defined as

$$H(\widehat{\beta}^{\mathrm{CF}}, \widehat{\beta}^{\mathrm{2SLS}}) = \left(\widehat{\beta}^{\mathrm{CF}} - \widehat{\beta}^{\mathrm{2SLS}}\right)^T \left[Cov\left(\widehat{\beta}^{\mathrm{2SLS}}\right) - Cov\left(\widehat{\beta}^{\mathrm{CF}}\right)\right]^- \left(\widehat{\beta}^{\mathrm{CF}} - \widehat{\beta}^{\mathrm{2SLS}}\right), \quad (23)$$

where $Cov\left(\widehat{\beta}^{\mathrm{CF}}\right)$ and $Cov\left(\widehat{\beta}^{\mathrm{2SLS}}\right)$ are the covariance matrices of $\widehat{\beta}^{\mathrm{CF}}$ and $\widehat{\beta}^{\mathrm{2SLS}}$ and $A^-$ denote the Moore-Penrose pseudoinverse. Under the null hypothesis where the augmented instrument $error.iv$ is valid, the test statistic $H(\widehat{\beta}^{\mathrm{CF}}, \widehat{\beta}^{\mathrm{2SLS}})$ is asymptotically $\chi_1^2$.

Based on the statistic $H(\widehat{\beta}^{\mathrm{CF}}, \widehat{\beta}^{\mathrm{2SLS}})$, we introduce a pretest estimator in Algorithm 3, which uses the control function estimator if there is not evidence that it is inconsistent and otherwise uses the usual two stage least squares estimator.

---

**Algorithm 3** Pretest estimator

**Input:** i.i.d observations of pre-treatment covariates $z_1$, instrumental variables $(z_2, h_2(z_2) \cdots, h_k(z_2))$, the treatment variable $y_2$, outcome variable $y_1$ and level $\alpha$.

**Output:** The pretest estimator of treatment effects $\beta = (\beta_1, \beta_2, \cdots, \beta_k)$ in (2).

1: Implement Algorithm 1 and obtain the estimator $\widehat{\beta}^{\mathrm{2SLS}}$ and the corresponding covariance structure $Cov\left(\widehat{\beta}^{\mathrm{2SLS}}\right)$. Implement Algorithm 2 and obtain the estimator $\widehat{\beta}^{\mathrm{CF}}$ and the corresponding covariance structure $Cov\left(\widehat{\beta}^{\mathrm{CF}}\right)$.

2: Calculate the test statistic $H(\widehat{\beta}^{\mathrm{CF}}, \widehat{\beta}^{\mathrm{2SLS}})$ as (23) and define the p-value $p = P\left(\chi^2(1) \geq H(\widehat{\beta}^{\mathrm{CF}}, \widehat{\beta}^{\mathrm{2SLS}})\right)$. The level $\alpha$ pretest estimator is defined as

$$\widehat{\beta}^{\mathrm{Pretest}} = \begin{cases} \widehat{\beta}^{\mathrm{CF}} & \text{if } p > \alpha \\ \widehat{\beta}^{\mathrm{2SLS}} & \text{if } p \leq \alpha . \end{cases} \quad (24)$$

---

For the rest of the paper, we will use $\alpha = 0.05$ in Algorithm 3. If the p-value of the test statistic $H(\widehat{\beta}^{\mathrm{CF}}, \widehat{\beta}^{\mathrm{2SLS}})$ is less than 0.05, then there is evidence that the control function estimator is inconsistent and the usual two stage least squares estimator is used; otherwise, the control function estimator is used.

## 5. Simulation study

Our simulation study compares the usual two stage least squares (2SLS) estimator, the control function (CF) estimator and the pretest estimator. We will consider several different model settings where the assumptions of the control function estimator are satisfied, moderately violated (two settings) and drastically violated and we will also investigate the sensitivity of the control function estimator to different joint distributions of errors. The sample size is 10,000 and we implement 10,000 simulations for each setting. We report the winsorized sample mean of the estimators (WMEAN) and the winsorized root mean square error (WRMSE). The non-winsorized sample mean of the estimators (NWEAN) and the non-winsorized root mean square error (NRMSE) are reported in the appendix C. The winsorized statistics are implemented as setting the 95-100 percentile as the 95-th percentile and the 0-5 percentile as the 5-th percentile and using the winsorized data to calculate the mean and root mean square error. The winsorized statistics are less sensitive to outliers (Wilcox & Keselman , 2003). The non-winsorized results summarized in the appendix C have similar patterns as the winsorized results reported in the following discussion. The outcome model that we are considering is

$$y_1 = \beta_0 + \beta_1 z_1 + \beta_2 y_2 + \beta_3 y_2^2 + u_1; \tag{25}$$

In the following subsections, we will generate $y_2$ according to different models.

### 5.1 Setting where the assumptions of the control function estimator are satisfied

For the setting where the assumptions of control function estimator are satisfied, the model we consider is:

$$y_1 = 1 + z_1 + 10y_2 + 10y_2^2 + u_1;$$
$$y_2 = 1 + \frac{1}{8}z_1 + \frac{1}{3}z_2 + \frac{1}{8}z_2^2 + v_2; \tag{26}$$

where $z_1 \sim N(0,1)$ and $z_2 \sim N(0,1)$ and $\begin{pmatrix} u_1 \\ v_2 \end{pmatrix} \sim N \left[ \begin{pmatrix} 0 \\ 0 \end{pmatrix}, \begin{pmatrix} 1 & 0.5 \\ 0.5 & 1 \end{pmatrix} \right]$. Setting (1) of Table 1 presents the absolute value of ratio of bias of the sample winsorized mean to the true value and the ratio of the sample winsorized root mean square error to the sample winsorized root mean square error of the two stage least squares estimator, where the columns under WMEAN present absolute value of the ratio of the bias and the columns under the WRMSE present the ratio of the root mean square error. Since the ratio of the sample winsorized root mean square error is using the sample winsorized root mean square error of the two stage least squares estimator as the basis, we only report the ratio of the ratio of the sample winsorized root mean square error for the control function estimator and the pretest estimator. ERR represents the empirical rejection rate of the null hypothesis that the extra instrumental variable given by the control function method is valid. The ERR is around 0.05 since the assumptions of CF estimator is satisfied in this model. The usual two stage least squares, control function and pretest estimator all have small bias. The WRMSE of control function estimator is consistently smaller than the WRMSE of the usual two stage least squares estimator. For $\beta_2$ and $\beta_3$, the WRMSE are more than 7 times smaller than that of the usual two stage least squares estimator. The pretest estimator also gains substantially over the two stage least squares estimator.

## 5.2 Setting where the assumptions of the control function estimator are moderately violated

We consider two models where the assumptions of the control function estimator are moderately violated,

1. $y_2$ is involved with a cubic function of $z_2$

$$y_1 = 1 + z_1 + 10y_2 + 10y_2^2 + u_1;$$
$$y_2 = \frac{1}{2} \times \frac{\left(1 + \frac{1}{8}z_1 + \frac{1}{3}z_2 + \frac{1}{8}z_2^2 + z_2^3\right)}{\operatorname{sd}\left(1 + \frac{1}{8}z_1 + \frac{1}{3}z_2 + \frac{1}{8}z_2^2 + z_2^3\right)} + v_2; \tag{27}$$

where $\operatorname{sd}\left(1 + \frac{1}{8}z_1 + \frac{1}{3}z_2 + \frac{1}{8}z_2^2 + z_2^3\right)$ is the standard deviation of $1 + \frac{1}{8}z_1 + \frac{1}{3}z_2 + \frac{1}{8}z_2^2 + z_2^3$ and $z_1 \sim N(0,1)$ and $z_2 \sim N(0,1)$ and $\begin{pmatrix} u_1 \\ v_2 \end{pmatrix} \sim N \left[ \begin{pmatrix} 0 \\ 0 \end{pmatrix}, \begin{pmatrix} 1 & 0.5 \\ 0.5 & 1 \end{pmatrix} \right]$.

2. $y_2$ is involved with an exponential function of $z_2$.

$$y_1 = 1 + z_1 + 10y_2 + 10y_2^2 + u_1$$
$$y_2 = \frac{1}{2} \times \frac{\left(1 + \frac{1}{8}z_1 + \frac{1}{3}z_2 + \frac{1}{8}z_2^2 + \exp(z_2)\right)}{\operatorname{sd}\left(1 + \frac{1}{8}z_1 + \frac{1}{3}z_2 + \frac{1}{8}z_2^2 + \exp(z_2)\right)} + v_2; \tag{28}$$

where $\operatorname{sd}\left(1 + \frac{1}{8}z_1 + \frac{1}{3}z_2 + \frac{1}{8}z_2^2 + \exp(z_2)\right)$ is the standard deviation of $1 + \frac{1}{8}z_1 + \frac{1}{3}z_2 + \frac{1}{8}z_2^2 + \exp(z_2)$ and $z_1 \sim N(0,1)$ and $z_2 \sim N(0,1)$ and $\begin{pmatrix} u_1 \\ v_2 \end{pmatrix} \sim N \left[ \begin{pmatrix} 0 \\ 0 \end{pmatrix}, \begin{pmatrix} 1 & 0.5 \\ 0.5 & 1 \end{pmatrix} \right]$.

In these simulation settings, since the terms $1 + \frac{1}{8}z_1 + \frac{1}{3}z_2 + \frac{1}{8}z_2^2 + z_2^3$ or $1 + \frac{1}{8}z_1 + \frac{1}{3}z_2 + \frac{1}{8}z_2^2 + \exp(z_2)$ involved in the $y_2$ model tend to be much larger than the error $v_2$, we standardize these by their standard deviation. The assumptions of the control function estimator are moderately violated in models (27) and (28) since we use a quadratic model of $z_2$ to fit the endogenous variable $y_2$, whose conditional mean has a cubic (exponential) term of $z_2$.

Setting (2) of Table 1 presents the absolute value of ratio of bias of the sample winsorized mean to the true value of the model (27) and the ratio of the sample winsorized root mean square error to the sample winsorized root mean square error of the two stage least squares estimator. The ratio of bias of two stage least squares, control function and pretest estimator to the true value are small. The empirical rejection rate is around 0.10 and the WRMSE of the control function estimator and the pretest estimator are smaller than WRSME of the two stage least squares estimator.

Setting (3) of Table 1 presents the WMEAN and WRMSE results for the exponential model (28). The ratio of the bias of two stage least squares, control function and pretest estimators to the true value are small. The empirical rejection rate is low and the WRMSE of the control function estimator and the pretest estimator are smaller than WRSME of the two stage least squares estimator. For $\beta_3$, the WRMSE of the CF estimator and the pretest estimator are around 5% of WRMSE of the usual two stage least squares estimator. Even though the assumption 3 of the control function method is violated for the models (27) and (28), the CF estimator is still approximately unbiased and more efficient than the 2SLS estimator. For these settings, the pretest estimator generally uses the CF estimator and is considerably more efficient than the 2SLS estimator.

### 5.3 Setting where the assumptions of the control function estimator are drastically violated

We now consider a setting in which the control function assumptions are drastically violated. The model setting is as follows:

$$
\begin{aligned}
y_2 &= -\gamma_2 + \gamma_1 z_2 + \gamma_2 z_2^2 + v_2 \\
w &= \delta v_2^2 + N(0,1) \\
y_1 &= \beta_2 y_2 + \beta_3 y_2^2 + \beta_4 w + u_1
\end{aligned}
\tag{29}
$$

where $z_2 \sim N(0,1)$ and $u_1, v_2 \sim N\left( \left( \begin{array}{c} 0 \\ 0 \end{array} \right), \left( \begin{array}{cc} 1 & 0 \\ 0 & 1 \end{array} \right) \right)$ and $\beta_2 = 1, \beta_3 = 0.2, \beta_4 = 1$ and $\gamma_1 = 1, \gamma_2 = 0.2$ and $\delta = 0.5$.

Setting (4) of Table 1 presents the absolute value of ratio of bias of the sample winsorized mean to the true value of the model (29) and the ratio of the sample winsorized root mean square error to the sample winsorized root mean square error of the two stage least squares estimator.

The ratio of bias of the WMEAN of usual two stage least squares and pretest estimators to the true value are close to zero while the ratio of the bias for the control function estimator is far away from zero. The WRMSE of the control function estimator is much larger than that of the usual two stage least squares estimator. The pretest estimator performs well in terms of sample mean and mean square error because the pretest hypothesis testing rejects most of the control function estimators.

### 5.4 Sensitivity of joint distribution of errors

In this section, we investigate how sensitive the control function estimator is to different joint distributions of $(u_1, v_2)$; in particular, we consider settings where Assumption 3 is satisfied but Assumption 4 is not satisfied. Assume that we use a quadratic model in $z_2$ to fit the first stage and the true first stage model is quadratic in $z_2$,

$$
\begin{aligned}
y_1 &= 1 + z_1 + 10y_2 + 10y_2^2 + u_1; \\
y_2 &= 1 + \frac{1}{8}z_1 + \frac{1}{3}z_2 + \frac{1}{8}z_2^2 + v_2;
\end{aligned}
\tag{30}
$$

where $z_1 \sim N(0,1)$ and $z_2 \sim N(0,1)$, but $u_1$ and $v_2$ are not bivariate normal. We generate $u_1$ and $v_2$ in the following three ways

| | WMEAN | | | WRMSE | |
|---|---|---|---|---|---|
| | 2SLS | Control Function | Pretest | Control Function | Pretest |
| (1) | | | Satisfied Model (26)(ERR=0.0510) | | |
| $\beta_2$ | 0.001 | 0.000 | 0.000 | 0.139 | 0.155 |
| $\beta_3$ | 0.000 | 0.000 | 0.000 | 0.070 | 0.079 |
| (2) | | | Cubic Model (27)(ERR=0.1073) | | |
| $\beta_2$ | 0.000 | 0.000 | 0.000 | 0.994 | 0.996 |
| $\beta_3$ | 0.000 | 0.000 | 0.000 | 0.635 | 0.736 |
| (3) | | | Exp Model (28) (ERR=0.0515) | | |
| $\beta_2$ | 0.001 | 0.001 | 0.000 | 0.275 | 0.311 |
| $\beta_3$ | 0.001 | 0.000 | 0.000 | 0.049 | 0.053 |
| (4) | | | Drastically Violated Model (29)(ERR=1.0000) | | |
| $\beta_2$ | 0.000 | 0.128 | 0.000 | 6.900 | 1.000 |
| $\beta_3$ | 0.000 | 0.546 | 0.000 | 10.275 | 1.000 |

Table 1: The ratio of the bias of the sample winsorized mean (WMEAN) to the true value and the ratio of the Winsorized RMSE (WRMSE) of the control function(Pretest) estimator to WRMSE of the two stage least squares estimator for Satisfied Model (26), Cubic Model (27) with SD = 0.5, Exponential Model (28) with SD = 0.5 and Drastically Violated Model (29). The Empirical Rejection Rate (ERR) stands for the proportion (out of 10,000 simulations) of rejection of the null hypothesis that the extra instrumental variable given by the control function method is valid.

1. Double exponential distribution: $u_1 = \epsilon_1$ and $v_2 = \frac{1}{2}\epsilon_1 + \frac{\sqrt{3}}{2}\epsilon_2$ where $\epsilon_1$ and $\epsilon_2$ are independent double exponential distribution with mean 0 and variance 1.

2. Bivariate log-normal distribution: $u_1 = \exp(\epsilon_1)$ and $v_2 = \exp(\epsilon_2)$ where $\epsilon_1, \epsilon_2 \sim N\left( \left( \begin{array}{c} 0 \\ 0 \end{array} \right), \left( \begin{array}{cc} 1 & 0.5 \\ 0.5 & 1 \end{array} \right) \right)$.

3. Bivariate absolutely normal distribution: $u_1 = |\epsilon_1| - E|\epsilon_1|$ and $v_2 = |\epsilon_2| - E|\epsilon_2|$ where $\epsilon_1, \epsilon_2 \sim N\left( \left( \begin{array}{c} 0 \\ 0 \end{array} \right), \left( \begin{array}{cc} 1 & 0.5 \\ 0.5 & 1 \end{array} \right) \right)$.

For these settings, Assumption 3 is satisfied but Assumption 4 is not satisfied. Table 2 presents the ratio of the bias of the sample winsorized mean (WMEAN) to the true value and the ratio of the Winsorized RMSE (WRMSE) of the control function(Pretest) estimator to WRMSE of the two stage least squares estimator for the model (30) and different joint distributions of errors. For the double exponential distribution and the log normal distribution, the ratio of the bias of the usual two stage least squares, the control function and the pretest estimator are small; the WRMSE of the control function and the pretest estimator are much smaller than that of the usual two stage least squares estimator. For the absolute normal distribution, the ratio of the bias of the control function and the pretest estimator are slightly larger than that of the usual two stage least squares; the WRMSE of the control function estimator is slightly smaller than that of the usual two stage least squares estimator and that of the pretest estimator is slightly larger than that of the usual two stage least squares estimator. In summary, the control function estimator is robust to the violations of Assumption 4 considered in this section.

| | WMEAN | | | WRMSE | |
|---|---|---|---|---|---|
| | 2SLS | Control Function | Pretest | Control Function | Pretest |
| (1) | Double exponential distribution (ERR=0.0520) | | | | |
| $\beta_2$ | 0.001 | 0.000 | 0.000 | 0.154 | 0.174 |
| $\beta_3$ | 0.000 | 0.000 | 0.000 | 0.099 | 0.112 |
| (2) | Log normal distribution(ERR=0.0509) | | | | |
| $\beta_2$ | 0.006 | 0.006 | 0.006 | 0.077 | 0.079 |
| $\beta_3$ | 0.001 | 0.001 | 0.001 | 0.060 | 0.059 |
| (3) | Absolute normal distribution(ERR=0.1455) | | | | |
| $\beta_2$ | 0.000 | 0.017 | 0.011 | 0.964 | 1.063 |
| $\beta_3$ | 0.000 | 0.004 | 0.002 | 0.968 | 1.065 |

Table 2: The ratio of the bias of the sample winsorized mean (WMEAN) to the true value and the ratio of the Winsorized RMSE (WRMSE) of the control function(Pretest) estimator to WRMSE of the two stage least squares estimator for the model (30) and different joint distributions of errors. For this model, the assumption 4 of the control function estimator to be consistent is violated.

## 6. Application to a field experiment in Burundi

In this section, we apply the control function method and usual two stage least squares to estimate the effect of exposure to violence on a person's patience (conflict on time preference). The data set consists of 302 observations from a field experiment in Burundi (Voors et al (2012)). Burundi underwent a civil war from 1993-2005, where the intensity of the conflict varied in different parts of the country. Voors et al (2012) sought to access whether the exposure to violence that a person experienced, as measured by the percentage dead in attacks the person lived ($y_{i2}$) affected a person's patience ($y_{i1}$), as measured by a person's discount rate for willingness to receive larger amounts of money in the future compared to smaller amounts of money now. A smaller value of the discount rate $y_{i1}$ means the person is more patient. The exogenous variables $\mathbf{x}_i$ that are available are whether the respondent is literate, the respondent's age, the respondent's sex, the total land holding per capita, land Gini coefficient, distance to market, conflict over land, ethnic homogeneity, socioeconomic homogeneity, population density and per capita total expenditure. As discussed in Voors et al (2012), it is possible that the exposure to violence ($y_{i2}$) is endogenous because violence may be targeted in a non-random way that is related to patience of the community; for example, violence may be targeted to extract "economic profit", i.e., steal the assets of others, where communities which are more vulnerable to having their assets stolen may also differ in their patience levels from less vulnerable communities. Due to possibility of endogeneity, the OLS estimator of regressing the outcome on the observed covariates maybe a biased estimate of the effect of exposure to violence on time preference. We follow the discussion in the paper Voors et al (2012) and use the following instrumental variables, distance to Bujumbura (the capital of Burundi) and altitude. Fighting was more intense near Bujumbura (the capital of Burundi) and at higher altitudes. The assumption for distance to Bujumbura and altitude to be valid instruments is that they only affect the distribution of violence and are not associated with the patience level of a community. Voors et al (2012) defend this assumption by discussing that most plausible concern with its validity is that distance to the capital and altitude might be associated with distance to market which might be associated with patience, but that in Burundi, since most farmers operate at a substance level, selling goods to nearby market, there are local markets in all communities, reducing the possibility of geography being correlated with preference. See more detailed discussion in page 956 in Voors et al (2012). We are able to express our model as

$$y_{1i} = \beta_0 + \beta_1^T \mathbf{x}_i + \beta_2 y_{2i} + \beta_3 y_{2i}^2 + u,$$

and

$$y_{2i} = \alpha_0 + \alpha_1 z_{i1} + \alpha_2 z_{i2} + \alpha_3 z_{i1}^2 + \alpha_4 z_{i2}^2 + \alpha_5 z_{i1} \times z_{i2} + \alpha_6^\mathsf{T} \mathbf{x}_i + v,$$

where $z_{i1}$ represents the distance to Bujumbura and $z_{i2}$ represents the altitude for the $i$-th subject.

|         | two stage least squares | control function |
|---------|:-----------------------:|:----------------:|
| $\beta_2$ | -1.282                | 2.101            |
| (SD)    | (7.033)                 | (3.479)          |
| $\beta_3$ | 0.2360                | 0.017            |
| (SD)    | (0.451)                 | (0.217)          |

Table 3: Estimates and Standard Error of Coefficients before Endogenous Variables

Table 3 presents the control function estimate and two stage least squares estimate of $\beta_2$ and $\beta_3$. Rather than focus directly on $\beta_2$ and $\beta_3$, we focus on a more interpretable quantity, the increase of discount rate when percentage dead in attacks increase from $y_2$ to $y_2 + 1$, which we denote by $\delta(y_2)$,

$$\delta(y_2) = y_1^{y_2+1} - y_1^{y_2} = \beta_2 + \beta_3 (1 + 2y_2).$$

In Figure 3(a)/(b), we plot usual two stage least squares/control function point estimates and corresponding confidence intervals of $\delta(y_2)$ respectively. The middle curve represents the point estimate of $\delta(y_2)$, and the higher curve and lower curve represent the upper and lower bound of 95% confidence interval of $\delta(y_2)$ respectively. In Figure 3(c), we compare the variance of $\delta(y_2)$ provided by control function and usual two stage least squares by plotting the ratio:

$$\frac{Var.2SLS\left(\delta(y_2)\right)}{Var.CF\left(\delta(y_2)\right)},$$

where $Var.2SLS\left(\delta(y_2)\right)$ represents the variance of $\delta(y_2)$ given by usual two stage least squares and $Var.CF\left(\delta(y_2)\right)$ represents the variance of $\delta(y_2)$ given by control function. Figure 3(c) illustrates that the ratio of variance is always larger than one and shows that the control function estimate is more efficient than the usual two stage least squares estimate. We apply the Hausman test to this real data and the corresponding pvalue is around 0.579 so that the validity of the extra instrumental variable is not rejected and the pretest estimator is equal to the control function estimator.

Figure 3(d) demonstrates the predicted causal relationship between the discount rate and percentage dead in attacks. All the other covariates are set at the sample mean level and we plug in the estimates of the coefficients by the control function method and the two stage least square method. As illustrated, we see that the exposure to violence (measured by percentage dead in attacks) decreases patience (measured by a smaller discount rate).

We present two additional applications of the control function method to estimate the effect of household income on food demand in Appendix D and the effect of smoking during pregnancy on birthweight on Appendix E.

## 7. Discussion

The model (2) assumes constant treatment effect. The comparison between the CF estimator and the 2SLS estimator presented in this paper can be extended to more general class of models that allows for heterogeneous treatment effects model, as discussed in section 5 of Small (2007). Without loss of generality, we consider the outcome model to be a quadratic model.

$$y_{1,i} = \beta_0 + \beta_{1,i} y_{2,i} + \beta_{2,i} y_{2,i}^2 + \beta_3 z_{1,i} + u_{1,i}, \quad \text{for} \quad i = 1, \cdots, n, \tag{31}$$

and assume that $z_{2,i}$ and $z_{2,i}^2$ are valid instruments for $y_{2,i}$ and $y_{2,i}^2$. We introduce the following notation $\beta_1 = \mathbb{E}\beta_{1,i}$ and $\beta_2 = \mathbb{E}\beta_{2,i}$ and assume that

$$(\beta_{1,i}, \beta_{2,i}) \quad \text{is independent of} \quad y_{2,i}, z_{1,i}, z_{2,i}, u_{1,i}, \tag{32}$$

which is a special case of assumption A.3 in Wooldridge (1997) by setting $\rho_1 = 0$. As discussed in Wooldridge (1997) and Small (2007), the assumption (32) states that the units do not select their treatment levels ($y_{2,i}$ and $y_{2,i}^2$) based on the gains that they would experience from the treatment ($\beta_{1,i}$ and $\beta_{2,i}$). The heterogenous effect model (31) can be expressed as

$$y_{1,i} = \beta_0 + \beta_1 y_{2,i} + \beta_2 y_{2,i}^2 + \beta_3 z_{1,i} + \left(u_{1,i} + (\beta_{1,i} - \beta_1) y_{2,i} + (\beta_{2,i} - \beta_2) y_{2,i}^2\right), \quad \text{for} \quad i = 1, \cdots, n.$$

The new error $\left(u_{1,i} + (\beta_{1,i} - \beta_1) y_{2,i} + (\beta_{2,i} - \beta_2) y_{2,i}^2\right)$ has the following property

$$\mathbb{E}\left(u_{1,i} + (\beta_{1,i} - \beta_1) y_{2,i} + (\beta_{2,i} - \beta_2) y_{2,i}^2\right) z_{1,i} = \mathbb{E}\left(u_{1,i} z_{1,i}\right) + \mathbb{E}\left(\beta_{1,i} - \beta_1\right) y_{2,i} z_{1,i} + \mathbb{E}\left(\beta_{2,i} - \beta_2\right) y_{2,i}^2 z_{1,i} = 0$$

where the first part follows from that $z_{1,i}$ is exogenous and the last two part follows from the assumption (32) . Similarly, we can also show that

$$\mathbb{E}\left(\left(u_{1,i} + (\beta_{1,i} - \beta_1) y_{2,i} + (\beta_{2,i} - \beta_2) y_{2,i}^2\right) z_{2,i}\right) = 0,$$

and

$$\mathbb{E}\left(\left(u_{1,i} + (\beta_{1,i} - \beta_1) y_{2,i} + (\beta_{2,i} - \beta_2) y_{2,i}^2\right) z_{2,i}^2\right) = 0.$$

Hence, the instrumental variables are valid for the new error. Our theory for the homogeneous model (2) can be extended to the general (31) under the assumption (32) (it can also be extended under the weaker assumptions than (32) found in Wooldridge (1997)). The pretest estimator provides a consistent estimator and is close to the more efficient estimator between 2SLS and the CF estimator.

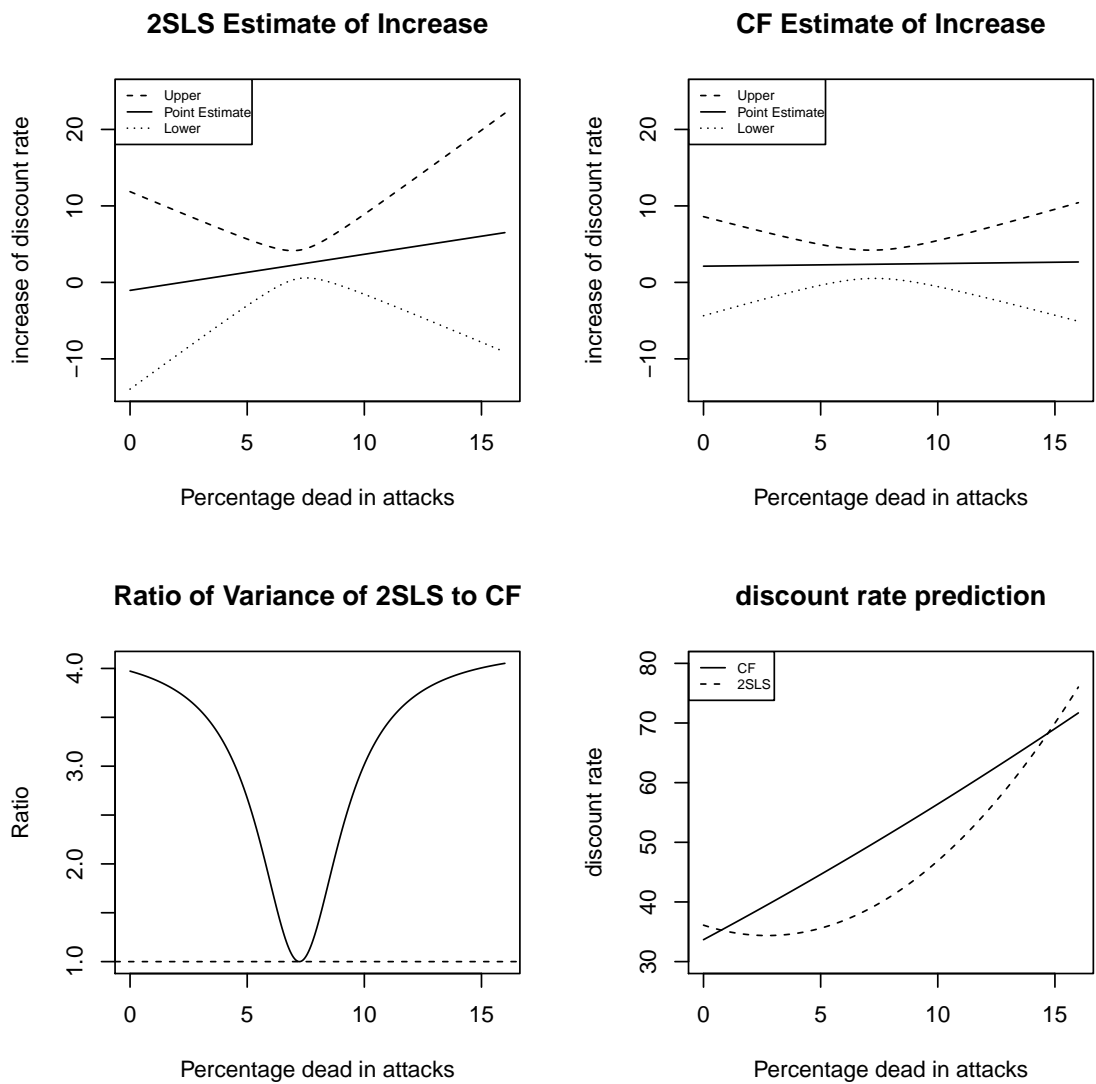

Figure 3: (a) plots the two stage least squares estimate and 95% confidence interval for the increase of discount rate $\delta(y_2)$ with respect to different percentage dead in attacks;(b) plots the control function estimate and 95% confidence interval for the increase of discount rate $\delta(y_2)$ with respect to different percentage dead in attacks;(c) plots the ratio of the variance of two stage least squares to the variance of the control function method with respect to different percentage dead in attacks;(d) plots the discount rate as a function of the percentage dead in attacks with the CF estimate and the 2SLS estimate.

We focus on the continuous outcome model in this paper. For a count outcome that follows a log linear model,

$$\log \mathbb{E}\left(y_1 | y_2, z\right) = \beta_0 + \beta_1 y_2 + \beta_2 z_1 + \beta_3 u. \tag{33}$$

where

$$y = \alpha_0 + \alpha_1 z_1 + \alpha_2 z_2 + \delta,$$

and $z_1$ is the included exogenous variable and $z_2$ is the instrumental variable and $\delta$ is normal variable with zero mean and variance $\sigma^2$, Mullahy (1997) shows that the 2SLS estimator is inconsistent while the CF estimator is consistent in this count outcome model. For a binary outcome that follows a logistic regression model, comparison of the CF estimator and the 2SLS estimator for the logistic second stage model can be a further research topic, discussion is found in Cai et al. (2011) and Clarke & Windmeijer (2012).

Imbens & Rubin (1997) formulated the potential outcome model for binary instrumental variable and binary treatment, which can be extended to continuous instrumental variable and treatment as follows: The treatment $y_2 = f_c\left(z\right)$ is an increasing function in $z$ given the compliance class $c$, where $c \in \{1, \cdots, M\}$. We also assume the monotonicity assumption of the compliance class

$$f_{c'}\left(z\right) \leq f_c\left(z\right) \quad \text{for} \quad c' \leq c.$$

The outcome model is

$$y_1 | c, z \sim N(f_c(z) + \lambda_c, \sigma^2).$$

As discussed in Imbens & Rubin (1997), the MLE and Bayesian estimator is more efficient than the instrumental variable based method since the instrumental variable method does not take into account the non-negativity of the density function. However, the estimator might be inconsistent if the model is wrong. In this paper, we focus on the semi-parametric model (2), which is robust to the outcome distribution. In the semi-parametric model (2), we show that the CF estimator is equivalent to a 2SLS estimator with an augmented set of instrumental variables. When the extra instrumental variables are valid, we demonstrate by theoretical results and simulation results that the CF estimator is more efficient than the usual 2SLS estimator.

## 8. Conclusion

This paper compares the control function estimator (Algorithm 2) and the usual two stage least squares estimator (Algorithm 1) for nonlinear causal effect models. Theoretically, we show that the control function estimator is equivalent to a two stage least squares estimator with an augmented set of instrumental variables. When the augmented instrumental variables are valid, the control function method will have a larger set of instrumental variables than the usual two stage least squares estimator and hence the control function estimator will have a smaller variance and root mean square error.

Methodologically, we develop a pretest estimator (Algorithm 3) which tests the validity of the augmented instrumental variables and combines the strength of the control function estimator and the usual two sage least squares estimator. In the simulation study, the pretest estimator provides large gains over two stage least squares in some settings while never being much worse than two stage least squares. Such phenomenon is also observed in the real data analysis.

For practical application, we suggest to use the pretest estimator which means implementing both the two stage least squares estimator and the control function estimator and calculating the Hausman test statistics based on these two estimates. If the Hausman test leads to the rejection of the null, the two stage least squares estimator is more trustworthy; otherwise, we use the control function estimator because it is more efficient and there is no evidence of it being biased.

## Appendix A. Consistency of the control function estimator

We now investigate why the control function method works under assumptions 1,2,3 and 4. Define $e = u_1 - \rho v_2$. Since $e$ is the residual of regressing $u_1$ on $v_2$, $e$ and $v_2$ are uncorrelated. By plugging in $u_1 = e + \rho v_2$, we have

$$y_1 = \beta_0 + \beta_1 y_2 + \cdots + \beta_k g_k(y_2) + \beta_{k+1} z_1 + \rho v_2 + e. \qquad (34)$$

Since $e$ is uncorrelated with $z_1$ and $v_2$, it suffices to show that the new error $e$ is uncorrelated with $g_i(y_2)$. By the law of iterated expectation and the one-to-one correspondence between $y_2$ and $v_2$, $E(g_i(y_2)e) = E\left[E(g_i(y_2)e|y_2,z)\right] = E\left[g_i(y_2)E(e|v_2,z)\right]$. By Assumption 3, $E(e|v_2,z) = E(u_1 - \rho v_2|v_2, z) = E(u_1 - \rho v_2|v_2)$. By Assumption 4, $E(u_1 - \rho v_2|v_2) = E(u_1|v_2) - \rho v_2 = 0$. Hence, $E(g_i(y_2)e) = 0$. and the new error $e$ is uncorrelated with $g_i(y_2)$. If we estimate $v_2$ by its estimate $\hat{v}_2$, the residual of the first stage $y_2 \sim z_1 + z_2 + h_2(z_2) + \cdots + h_k(z_2)$, the consistency of the control function estimate follows if we can verify the regularity conditions in Theorem 1 in Murphy et al (2002), which says that if we obtain the consistent estimates of first stage coefficients, we estimate the second stage coefficients consistently by replacing the residual with its estimate from the first stage. Let $Z$ denote the data matrix where each row is an i.i.d sample of $\{z_1, z_2, h_2(z_2), \cdots, h_k(z_2)\}$, $X$ denote the data matrix where each row is an i.i.d sample of $\{z_1, y_2, g_2(y_2), \cdots, g_k(y_2)\}$ and $V_2$ denote the data matrix where each row is an i.i.d sample of $\{\hat{v}_2\}$. The regularity conditions for Theorem 1 in Murphy et al (2002) are the following:

(a) $\lim_{n \to \infty} A^T A = Q_0$ where $A = (X, V_2)$ and $Q_0$ is positive definite matrix.

(b) $v_2 = f(\alpha, Z) = y_2 - (\alpha_0 + \alpha_1 z_1 + \alpha_2 \mathbf{H}(z_2))$ is twice continuously differentiable in $\alpha$ for each $Z$ and $\lim_{n \to \infty} A^T Z = Q_1$.

(c) The first stage estimate $\hat{\alpha}$ is a consistent estimator of $\alpha$.

We will show that under the following three regularity assumptions, the conditions $(a), (b), (c)$ of Theorem 1 in Murphy et al (2002) are satisfied.

  i $\lim_{n \to \infty} A^T A = Q_0$ where $A = (X, V_2)$ and $Q_0$ is positive definite matrix.

  ii $\lim_{n \to \infty} A^T Z = Q_1$.

  iii The true model of the first stage is (4).

Assumption i is the same as condition (a). Assumption ii is the second part of condition (b). For the first part of condition (b), since $v_2 = f(\alpha, Z)$ is a linear function of the parameters $\alpha$, $f$ is twice continuously differentiable. For condition (c), by assuming iii that the true model of the first stage is (4), the first stage regression coefficients $\hat{\alpha}$ of the standard control function estimator (7) are consistent estimates of $\alpha$.

## Appendix B. Proof of the theorems

**Lemma 7** (*Adjustment Lemma*)
*For the linear model*

$$y = \beta_1 x_1 + \beta_2 x_2 + \epsilon,$$

*the following two regression give us the same estimate of $\beta_1$:*

$$y \sim x_1 + x_2,$$

$$y \sim x_{1.2};$$

*where $x_{1.2}$ is the residual of the regression $x_1 \sim x_2$.*

This lemma is called Adjustment Lemma since we adjust $x_1$ to $x_2$ to obtain $x_{1.2}$.

**Proof** [Proof of Theorem 1] By the adjustment lemma, the coefficients of the regression $y_1 \sim \widetilde{z_1} + \widetilde{y_2} + \widetilde{g_2(y_2)}$ are the same with corresponding coefficients in $y_1 \sim z_1 + y_2 + g_2(y_2) + e_1$ where $\widetilde{x}$ is the residual of the regression $x \sim e_1$. Since $z_1$ is orthogonal to $e_1$ by the property of OLS, $\widetilde{z_1} = z_1$. The theorem follows by replacing $\widetilde{z_1}$ with $z_1$. ∎

**Proof** [Proof of Theorem 2] Let $Z = (1, z_1, \widetilde{z_2, h_2(z_2)})$ denote the matrix of observed values of instrumental variables. $y_2^2 = \gamma_c + \gamma e_1 + \widetilde{g_2(y_2)}$ Since $e_1$ is the residual of the regression

$$y_2 \sim z_1 + z_2 + h_2(z_2),$$

it follows from the property of OLS that $Z^T e_1 = 0_{4\times 1}$. The coefficients given by regression $g_2(y_2) \sim z_1 + z_2 + h_2(z_2)$ are $\left(Z^T Z\right)^{-1} Z^T g_2(y_2)$ and those given by regression $\widetilde{g_2(y_2)} \sim z_1 + z_2 + h_2(z_2)$ are $\left(Z^T Z\right)^{-1} Z^T \widetilde{g_2(y_2)}$. Since

$$\left(Z^T Z\right)^{-1} Z^T g_2(y_2) - \left(Z^T Z\right)^{-1} Z^T \widetilde{g_2(y_2)} = \left(Z^T Z\right)^{-1} Z^T e_1 = 0$$

we come to the conclusion that the regression $g_2(y_2) \sim z_1 + z_2 + h_2(z_2)$ and $\widetilde{g_2(y_2)} \sim z_1 + z_2 + h_2(z_2)$ give the same estimates of coefficients on $1, z_1, z_2, h_2(z_2)$. ∎

**Proof** [Proof of Theorem 3] It is sufficient to show that except for a constant difference,

(1) $\widetilde{y_2}$ is equal to the predicted value of the regression $y_2 \sim z_1 + z_2 + h_2(z_2) + error.iv.$

(2) $\widetilde{g_2(y_2)}$ is equal to the predicted value of the regression $g_2(y_2) \sim z_1 + z_2 + h_2(z_2) + error.iv.$

By the discussion after theorem 2, we have the following decomposition

$$g_2(y_2) = \gamma_c + \gamma e_1 + \overbrace{g_2(y_2)}$$
$$= \gamma_c + \gamma e_1 + \gamma_1 z_1 + \gamma_2 z_2 + \gamma_3 h_2(z_2) + error.iv.$$

(35)

where the column vector $e_1$ is orthogonal to the column vectors $z_2, h_2(z_2)$ and $error.iv$ and $error.iv$ is orthogonal to $z_1, z_2$ and $h_2(z_2)$ by the property of OLS. By the decomposition (35), we conclude that except for a constant, $\overbrace{g_2(y_2)}$ is equal to the predicted value of the regression $g_2(y_2) \sim z_1 + z_2 + h_2(z_2) + error.iv$. Since $error.iv$ is orthogonal to $z_1, z_2, h_2(z_2)$ and $e_1$, $error.iv$ is orthogonal to $y_2 = \alpha_0 + \alpha_1 z_1 + \alpha_2 z_2 + \alpha_3 h_2(z_2) + e_1$ and hence the predicted value of the regression $y_2 \sim z_1 + z_2 + h_2(z_2) + error.iv$ is the same with $\widehat{y_2}$, which is the same with $\widetilde{y_2}$. ∎

**Proof** [Proof of Theorem 4]

It is sufficient to show that except for a constant difference,

(1) $\widetilde{y_2}$ is equal to the predicted value of the regression

$$y_2 \sim z_1 + z_2 + h_2(z_2) + h_3(z_2) + error.iv_1 + error.iv_2.$$

(2) $\overbrace{g_2(y_2)}$ is equal to the predicted value of the regression

$$g_2(y_2) \sim z_1 + z_2 + h_2(z_2) + h_3(z_2) + error.iv_1 + error.iv_2.$$

(3) $\overbrace{g_3(y_2)}$ is equal to the predicted value of the regression

$$g_3(y_2) \sim z_1 + z_2 + h_2(z_2) + h_3(z_2) + error.iv_1 + error.iv_2.$$

We start with a decomposition of $g_3(y_2)$

$$g_3(y_2) = \gamma_c + \gamma e_1 + \overbrace{g_3(y_2)}$$
$$= \gamma_c + \gamma e_1 + \gamma_1 z_1 + \gamma_2 z_2 + \gamma_3 h_2(z_2) + \gamma_4 h_3(z_2) + error.iv_2.prime$$
$$= \gamma_c + \gamma e_1 + \gamma_1 z_1 + \gamma_2 z_2 + \gamma_3 h_2(z_2) + \gamma_4 h_3(z_2) + \gamma' error.iv_1 + error.iv_2.$$

(36)

Since $e_1$ is orthogonal to $error.iv_1, error.iv_2$ and $z_1, z_2, h_2(z_2), h_3(z_2)$, the predicted value given by the regression $g_3(y_2) \sim z_1 + z_2 + h_2(z_2) + h_3(z_2) + error.iv_1 + error.iv_2$ is the same with $\overbrace{g_3(y_2)}$.

Since $error.iv_2$ is orthogonal to $e_1, z_1, z_2, h_2(z_2), h_3(z_2), error.iv_1$, $error.iv_2$ is orthogonal to $g_2(y_2)$, hence the coefficient of $error.iv_2$ given by the following regression

$$g_2(y_2) \sim z_1 + z_2 + h_2(z_2) + h_3(z_2) + error.iv_1 + error.iv_2$$

(37)

is vanishing, which leads to the predicted value of regression (37) is the same with the predicted value of $g_2(y_2) \sim z_1 + z_2 + h_2(z_2) + h_3(z_2) + error.iv_1$ and by the decomposition (35) and same argument in proof of theorem 3, we obtain (2).

| | | NMEAN | | NRMSE | |
|---|---|---|---|---|---|
| | 2SLS | control function | Pretest | control function | Pretest |
| (1) | | Satisfied Model (26)(ERR=0.0510) | | | |
| $\beta_2$ | 0.001 | 0.000 | 0.000 | 0.136 | 0.543 |
| $\beta_3$ | 0.000 | 0.000 | 0.000 | 0.069 | 0.535 |
| (2) | | Cubic Model (27) (ERR=0.1073) | | | |
| $\beta_2$ | 0.000 | 0.000 | 0.000 | 0.993 | 0.995 |
| $\beta_3$ | 0.000 | 0.000 | 0.000 | 0.619 | 0.808 |
| (3) | | Exp Model (28) (ERR=0.0515) | | | |
| $\beta_2$ | 0.002 | 0.000 | 0.000 | 0.236 | 0.602 |
| $\beta_3$ | 0.001 | 0.000 | 0.000 | 0.040 | 0.546 |
| (4) | | Drastically Violated Model (29)(ERR=1.0000) | | | |
| $\beta_2$ | 0.000 | 0.128 | 0.000 | 6.278 | 1.000 |
| $\beta_3$ | 0.000 | 0.546 | 0.000 | 9.357 | 1.000 |

Table 4: The ratio of the bias of the sample non-winsorized mean(NMEAN)to the true value and the ratio of the Non-winsorized RMSE(NRMSE) of the control function(Pretest) estimator to NRMSE of the two stage least squares estimator for Satisfied Model (26), Cubic Model (27) with SD = 0.5, Exponential Model (28) with SD = 0.5 and Drastically Violated Model (29). The Empirical Rejection Rate (ERR) stands for the proportion (out of 10,000 simulations) of rejection of the null hypothesis that the extra instrumental variable given by the control function method is valid.

Since $error.iv_1$ and $error.iv_2$ are orthogonal to $e_1, z_1, z_2, h_2(z_2), h_3(z_2)$, then $error.iv_1$ and $error.iv_2$ are orthogonal to $y_2$ and (1) follows. ∎

## Appendix C. Further results from the simulation study

All the corresponding non-winsorized results are presented in Tables 4 and 5. The observations are similar to the counter part of winsorized results.

## Appendix D. Application to demand for food

In this section, we apply the control function method and usual two stage least squares to estimate the effect of income on food demand and compare these two methods. The data set is from Bouis & Haddad (1990) and comes from a survey of farm households in the Bukidnon Province of Philippines. Following Bouis & Haddad (1990), we assume that food expenditure is a quadratic function of log income. Here, $y_{1i}$ is the expenditure on food of the i-th household, $y_{2i}$ is the log of income of the i-th household and $x_i$ represents the included exogenous variables of the i-th household, which consist of mother's education, father's education, mother's age, father's age, mother's nutritional knowledge, corn price, rice price , population density of the municipality, number of household members expressed in adult equivalents, and dummy variables for the round of the survey. Then we are able to express our model as $y_{1i} = \beta_0 + \beta_1 x_i + \beta_2 y_{2i} + \beta_3 y_{2i}^2 + u$. Bouis & Haddad (1990) were concerned that regression of $y_1$ on $y_2$ (log income) and $x$ would not provide an unbiased estimate of $\beta$ because of unmeasured confounding variables. In particular, because farm households make production and consumption decisions simultaneously and there are multiple incomplete markets in the study area, the households' production decisions (which affect their log income $y_2$) are associated with their preferences according to microeconomic theory (Bardhan & Udry (1999), chap. 2). To solve the problem, Bouis & Haddad (1990) proposed cultivated area per capita as an instrumental variable. Bouis and Haddad's reasoning for why cultivated area per capita is an instrumental variable is that "land availability is assumed to be a constraint in the short run, and therefore exogenous to the household decision making process."

| | WMEAN | | | WRMSE | |
|---|---|---|---|---|---|
| | 2SLS | Control Function | Pretest | Control Function | Pretest |
| (1) | Double exponential distribution(ERR=0.0520) | | | | |
| $\beta_2$ | 0.001 | 0.000 | 0.000 | 0.150 | 0.554 |
| $\beta_3$ | 0.000 | 0.000 | 0.000 | 0.096 | 0.546 |
| (2) | Log normal distribution(ERR=0.0509) | | | | |
| $\beta_2$ | 0.002 | 0.006 | 0.009 | 0.003 | 0.310 |
| $\beta_3$ | 0.000 | 0.001 | 0.001 | 0.002 | 0.299 |
| (3) | Absolute normal distribution(ERR=0.1455) | | | | |
| $\beta_2$ | 0.000 | 0.017 | 0.010 | 0.881 | 1.027 |
| $\beta_3$ | 0.000 | 0.004 | 0.002 | 0.883 | 1.029 |

Table 5: The ratio of the bias of the sample non-winsorized mean(NMEAN)to the true value and the ratio of the Non-winsorized RMSE(NRMSE) of the control function(Pretest) estimator to NRMSE of the two stage least squares estimator for the model (30) and different joint distributions of errors. For this model, the assumption 4 of the control function estimator to be consistent is violated.

| | two stage least squares | control function |
|---|---|---|
| $\beta_2$ | 39.993 | 7.389 |
| (SD) | ( 22.353) | (3.728) |
| $\beta_3$ | -2.397 | 1.592 |
| (SD) | (2.730) | (0.431) |

Table 6: Estimates and Standard Error of Coefficients before Endogenous Variables

Table 6 presents the control function estimate and two stage least squares estimate of $\beta_2$ and $\beta_3$. Rather than focus directly on $\beta_2$ and $\beta_3$, we focus on a more interpretable quantity, the income elasticity of food demand at the mean level of food expenditure. This is the percent change in food expenditure caused by a 1% increase in income for households currently spending at the mean food expenditure level, and we denote it by $\eta(y_2)$. The mean food expenditure of households is 31.14 pesos per capita per week and

$$\eta(y_2) = \frac{100}{31.14}\hat{\beta}_2 \log(1.01) + \hat{\beta}_3 \left( 2 \log(1.01) y_2 + \log(1.01)^2 \right).$$

In Figure 4(a), we plot OLS point estimates and corresponding confidence intervals of $\eta(y_2)$ with respect to different levels of log income $y_2$. The middle curve represents the point estimate of $\eta(y_2)$, and the higher curve and lower curve represent the upper and lower bound of 95% confidence interval of $\eta(y_2)$ respectively. In Figure 4(b)/(c), we plot usual two stage least squares/control function point estimates and corresponding confidence intervals of $\eta(y_2)$ respectively. In Figure 4(d), we compare the variance of $\eta(y_2)$ provided by control function and usual two stage least squares by plotting the ratio:

$$\frac{Var.2SLS\left(\eta(y_2)\right)}{Var.CF\left(\eta(y_2)\right)},$$

where $Var.2SLS\left(\eta(y_2)\right)$ represents the variance of $\eta(y_2)$ given by usual two stage least squares and $Var.CF\left(\eta(y_2)\right)$ represents the variance of $\eta(y_2)$ given by control function. Figure 4(b) shows that two stage least squares leads to negative estimate of the non-negative valued Income Elasticity of Food Demand; in contrast, Figure 4(c) shows that the control function estimate of the non-negative valued Income Elasticity of Food Demand is always positive. Figure 4(d) illustrates that the ratio of variance is always larger than one and shows that the control function estimate is more efficient than the usual two stage least squares estimate. We apply the Hausman test to this real data and the corresponding pvalue is around 0.14 so that the validity of the extra instrumental variables is not rejected and the pretest estimator is equal to the control function estimator.

## Appendix E. Application to smoking data

In this section, we apply the control function method and usual two stage least squares to estimate the effect of maternal smoking during pregnancy on birth weight. The data set consists of 1388 observations from the Child Health Supplement to the 1998 National Health Interview Survey. The goal is to estimate the effect of maternal smoking (during pregnancy) on the birth weight. For the $i$-th observation, the outcome $y_{i1}$ represents the log birth weight, $y_{i2}$ represents the typical number of cigarettes smoked per day during pregnancy and $\mathbf{x}_i$ represents the other exogenous variables: birth order, race and child's sex. As discussed in Wehby et al. (2011), it is possible that mothers who smoke during pregnancy self-select into smoking based on their preferences for health and risk taking and their perceptions of fetal health endowments. These factors, which are typically unobserved in available data samples, can affect the fetal health through other pathways besides smoking. For example, women smoking during pregnancy might adopt other unhealthy behaviors having adverse effects on the fetus (e.g., poorer nutrition or reduced prenatal care). Since some of the confounding variables that are related with both smoking and child health are unobserved, the OLS estimator of regressing the outcome on the observed covariates maybe a biased estimate of the effect of smoking on the birth weight. We follow the discussion in the paper Mullahy (1997) and use the following instrumental variables $\mathbf{z}_i$: maternal schooling, paternal schooling, family income and the per-pack state excise tax on the cigarettes. Then

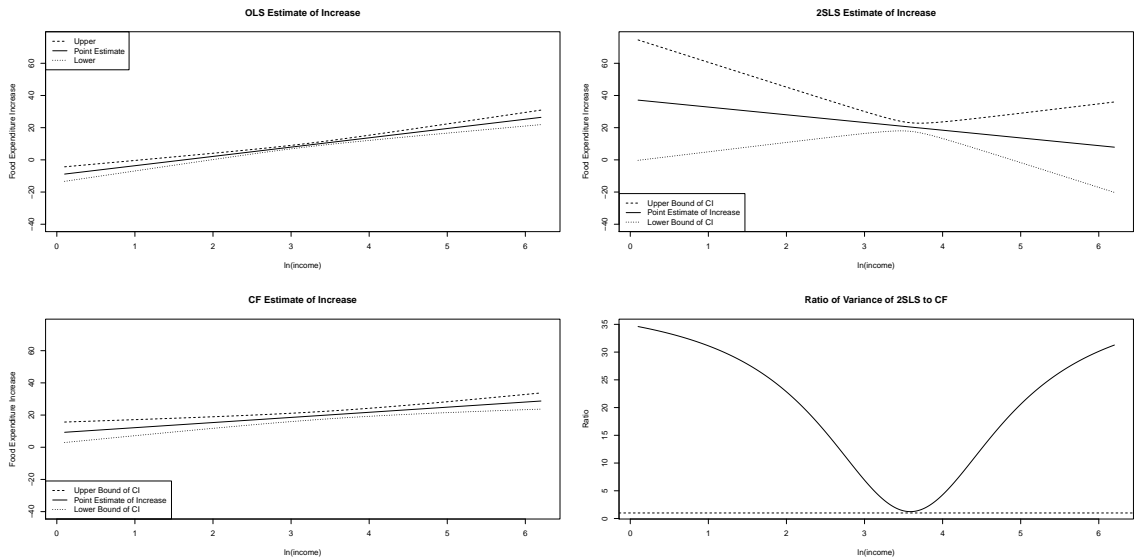

Figure 4: (a) plots the OLS estimate and 95% confidence interval for the income elasticity with respect to different log(income);(b) plots the two stage least squares estimate and 95% confidence interval for the income elasticity with respect to different log(income);(c) plots the control function estimate and 95% confidence interval for the income elasticity with respect to different log(income);(d) plots the ratio of the variance of two stage least squares to the variance of the control function method with respect to different log(income).

| | two stage least squares | control function |
|---|---|---|
| $\beta_2(\times 10^{-3})$ | -12.535 | -8.956 |
| $(\text{SD}(\times 10^{-3}))$ | (7.463) | (3.040) |
| $\beta_3(\times 10^{-3})$ | 0.266 | 0.121 |
| $(\text{SD}(\times 10^{-3}))$ | (0.284) | (0.068) |

Table 7: Estimates and Standard Error of Coefficients before Endogenous Variables

we are able to express our model as

$$y_{1i} = \beta_0 + \beta_1^T \mathbf{x}_i + \beta_2 y_{2i} + \beta_3 y_{2i}^2 + u,$$

and

$$y_{2i} = \alpha^T \mathbf{z}_i + v.$$

Table 7 presents the control function estimate and two stage least squares estimate of $\beta_2$ and $\beta_3$. Rather than focus directly on $\beta_2$ and $\beta_3$, we focus on a more interpretable quantity, the increase of log birthweight when the cigarettes increase from $y_2$ to $y_2 + 1$, which we denote by $\delta(y_2)$,

$$\delta(y_2) = y_1^{y_2+1} - y_1^{y_2} = \beta_2 + \beta_3 \left(1 + 2y_2\right).$$

In Figure 5(a)/(b), we plot usual two stage least squares/control function point estimates and corresponding confidence intervals of $\delta(y_2)$ respectively. The middle curve represents the point estimate of $\delta(y_2)$, and the higher curve and lower curve represent the upper and lower bound of 95% confidence interval of $\delta(y_2)$ respectively. In Figure 5(c), we compare the variance of $\delta(y_2)$ provided by control function and usual two stage least squares by plotting the ratio:

$$\frac{Var.2SLS\left(\delta(y_2)\right)}{Var.CF\left(\delta(y_2)\right)},$$

where $Var.2SLS\left(\delta(y_2)\right)$ represents the variance of $\delta(y_2)$ given by usual two stage least squares and $Var.CF\left(\delta(y_2)\right)$ represents the variance of $\delta(y_2)$ given by control function. Figure 5(c) illustrates that the ratio of variance is always larger than one and shows that the control function estimate is more efficient than the usual two stage least squares estimate. We apply the Hausman test to this real data and the corresponding pvalue is around 0.599 so that the validity of the extra instrumental variables is not rejected and the pretest estimator is equal to the control function estimator.

Figure 5(d) demonstrates the predicted causal relationship between the log birthweight and number of cigarettes smoked. All the other covariates are set at the sample mean level and we plug in the estimates of the coefficients by the control function method and the two stage least square method. As illustrated, we see that smoking during pregnancy has a negative effect on birthweight; with the number of cigarettes increasing, the absolute value of the marginal effect is decreasing, that is to say, the effect of additional cigarettes smoked becomes less.

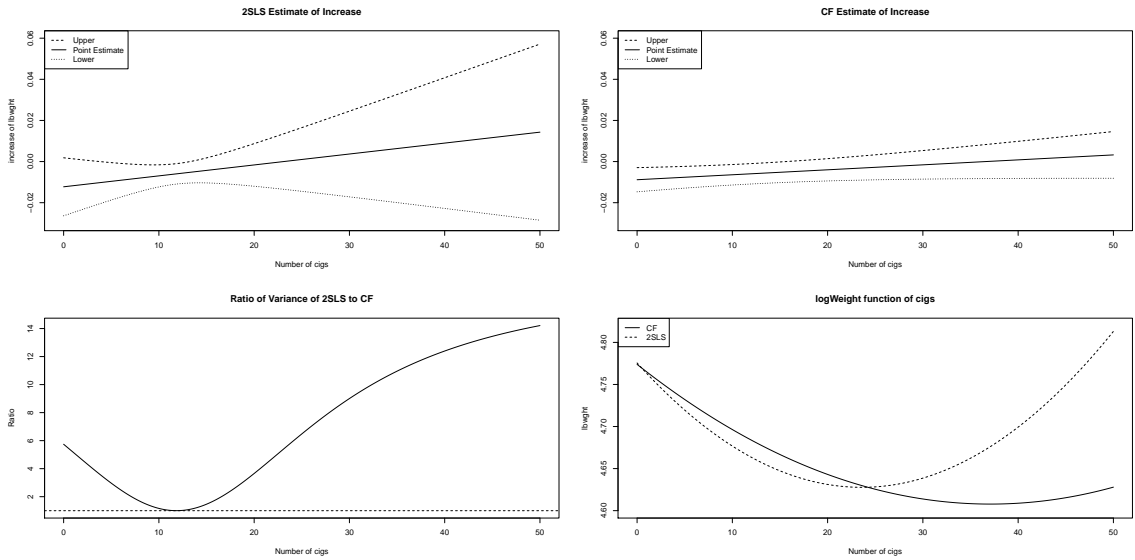

Figure 5: (a) plots the two stage least squares estimate and 95% confidence interval for the increase of lbwght $\delta(y_2)$ with respect to different number of cigarettes;(b) plots the control function estimate and 95% confidence interval for the increase of lbwght $\delta(y_2)$ with respect to different number of cigarettes;(c) plots the ratio of the variance of two stage least squares to the variance of the control function method with respect to different number of cigarettes;(d) plots the lbwght as a function of the number of cigarettes with the CF estimate and the 2SLS estimate.